\documentclass[epj]{svjour}   
\usepackage{graphics}
\usepackage{epsfig}
\usepackage{bm}

\begin{document}
\hbadness=10000
\title{$\Delta_{33}$-medium mass modification and pion spectra}
\author{Bi Pin-zhen\inst{1} \and Johann Rafelski\inst{2}}
\institute{Institute of Modern Physics, Fudan University, Shanghai 200433, China, \\ \email{pzbi@fudan.edu.cn}, \and
Department of Physics, University of Arizona, Tucson, Arizona, 85721, USA,\\  \email{rafelski@physics.arizona.edu}}
\date{\today}
\abstract{
We study the $\pi^\pm$-spectra obtained in 2, 4, 6 and  8$A$ GeV Au--Au collisions within the thermal model.  
We find that  the main features of the data can be well described after we include the pions from the decay of 
the $\Delta$-resonance with medium mass modification. 
}
\maketitle
\section{Introduction}  

Heavy ion collision experiments, in the energy range of several GeV,  
permit the study of nuclear  matter and hadron properties 
under  extreme conditions of temperature and density. Here, we are
interested in  how the pion spectra are influenced
by the  in medium mass modification. Considering  the 
most precise  pion spectra available, ,  obtained in the energy range 
2--8 $A$ GeV~\cite{Klay},  we show that there 
is sensitivity to the medium modification of the mass splitting 
between  $\Delta_{33}$ and $N$.  The availability of both 
pion charge polarities allows us  to address the pion isospin 
asymmetry, an important ingredient in our analysis. 

 A prominent  feature 
of these $\pi^\pm$-spectra   is that they are  not described   
by a single thermal  source. Instead, the initial analysis by the experimental
group used a two-component thermal model with two 
different source temperatures. 
The dynamical origin of such a  model is at present open to 
discussion. On the other hand, the decay   $\Delta_{33}(1232)\to N+\pi$ has  long 
been recognized to be a significant mechanism of pion production in  the
reaction energy range considered here~\cite{Hong:1997ka,Barrette:1994kq}.  
Moreover, the shape of  the decay pion spectrum  is different from the shape of a 
thermal spectrum~\cite{Brown:1991en}.  

Thus, in order to obtain the observed momentum spectra of pions,
we consider the  sum of the direct thermal pion component with 
the in-medium-decay  of the thermal $\Delta_{33}$
component. Only decays near the kinetic freeze-out
contribute in the second component: the early decay product 
pions are reequilibrated and are part of the pion thermal component.   
The number of particles at a given value 
of momentum (the spectrum) is established by considering statistical 
distributions which arise at temperature-dependent values of hadron masses.
Below the kinetic (i.e., scattering) 
freeze-out temperature  at about $T=100\pm20$  MeV,  in the free-streaming 
domain, the hadron mass values are restored to the free space value. Moreover,
as hadrons begin to free-stream out of the interaction region, their mass 
returns to the free space value, the required energy is derived 
from the modification of the volume-vacuum energy. 
Since all particles actually observed experimentally had time to `recover' from medium 
modifications to their properties, till now the experimental study of this medium effect
 has focused on the observation of the possible modification of the  
decay width in matter, especially of the $\phi$-meson~\cite{Rapp}.  

In order to describe the pion spectra we will show that 
 the required magnitude of the mass medium modification 
is in agreement with the widely held believes about the medium 
dependent properties of strongly interacting particles  in dense, 
hot matter~\cite{Brown:1991dk,Bro}. Considering 
 the quantum-hadrodynamic model~\cite{Serot:1997xg},
the cancellation of  several contributions to the energy of 
a nucleon with magnitude of a few hundred MeV
leads to  the relatively small
nuclear binding energy in normal nuclear matter. One of these
large components is   the scalar potential which modifies nucleon mass.
At normal nuclear density it has
a magnitude  of several 100 MeV, and as the nuclear matter is squeezed, 
this  potential increases such that the effective nucleon mass melts entirely
(see for example Fig 2. in Ref.\cite{Serot:1997xg}). Already at relatively moderate
temperatures the thermal melting is more important compared to the density effect.
A way to express the hybrid temperature -density dependence 
of mass modification is by the substitution:
\begin{equation}\label{bag2mu} 
T\to T_{\rm eff}\simeq \sqrt{T^2+(\mu_{\rm b}/2\pi)^2}. 
\end{equation}  
where $\mu_{\rm b}$ is baryochemical potential.

A brief description of the thermal 
model particle spectra is given in the following section \ref{therm-spec}.   
An analysis of the pion spectra  based on the proposed model with medium 
mass modification is presented in section \ref{analysis},   followed 
by a summary and discussion of our work in section~\ref{sum}.

\section{Particle spectra in the thermal model}\label{therm-spec} 
An important question governing the validity of the study  
presented here is: In what way is the pion spectrum  modified
by the medium?  When a medium-modified 
hadron emerges into medium free space, the hadron  mass relaxes 
and the hadron  picks up (or if appropriate releases) the   
energy from/to the vacuum, and  its mass returns to the normal free space value. 
In such a process, within an isotropic medium,  the translational momentum  
of a particle  cannot  change. However, along with the
mass, the energy   of the  particle will also  change. 
Since the momentum distribution of particles with $p=E/v$ 
is not changed in the process of free space mass restoration:\\
1) the number of thermal particles with  momentum $p$ is  
governed by the thermal momentum distribution of the source, obtained
with medium modified mass, 
for example in the Boltzmann (classical) limit:
\begin{equation}\label{thermBol}
{d^3N\over d^3p}\propto e^{-\sqrt{m(T)^2+p^2}/T},
\end{equation}
and thus for particle spectra all masses 
are to be understood to be medium-modified;\\
2) one should be able to  reconstruct a 
decaying resonance using 
the medium modified masses for both the resonance and its decay products, for example:
\begin{eqnarray}
m_\Delta^2 (T)  &=&  (E_\pi+E_{\rm N})^2\!-(\vec p_\pi+\vec p_{\rm N})^2    \nonumber\\[0.2cm]      
&=&
2\left(\sqrt{m^2_\pi(T)+p^2_\pi} \sqrt{m^2_{\rm N}(T)+p^2_{\rm N}}  -
     \vec p_\pi\cdot\vec p_{\rm N}\right)+    \nonumber\\[0.2cm]      
&+& m^2_\pi(T)+m^2_{\rm N}(T) .
\end{eqnarray} 
Here all momenta are as measured in the final state. 

This discussion shows that in order to obtain the observed momentum spectra of  
produced particles, we have to  use   medium-modified masses in the usual 
thermal model expressions, and discuss the results as function 
of the measured momentum. Other than the medium effect, 
our analysis proceeds along previously 
established methods~\cite{Sollfrank:1990qz,Sollfrank:1991xm,Letessier:1999rm}. The  thermal   
momentum distribution of  primary pions is, by assumption,
given  by the standard thermal Bose distribution (here $\hbar=c=k_B=1$), see  
also Eq.\,(\ref{thermBol}): 
\begin{eqnarray}\label{pispec1} 
{1 \over  2 \pi} {d\tilde N  \over  p_{\rm t} dp_{\rm t} dy }         
=N_\pi {{\Upsilon_\pi m_{\rm t} \cosh y  }          
\over {e^{\beta m_{\rm t} \cosh y } - \Upsilon_\pi}}. 
\end{eqnarray} 
${\beta}={1/T}$ and $N_{\pi}$ is the normalization factor, containing among other effects, the volume of  
the source. Rapidity $y=\tanh ^{-1} (p_{\rm l}/E)$ and transverse mass $m_{\rm t} =\sqrt{m^{2}+p_{\rm t}^2}$  
are convenient covariant variables. In the center of the hot, dense collision zone, high pressure is produced.  
This pressure will cause a collective motion of the system, matter flow \cite{Sollfrank:1990qz,Sollfrank:1991xm}.
Thus the parameter $T$ comprises, in a qualitative manner, the effect of radial matter flow. 
Using in  first approximation the Doppler formula, 
\begin{equation}T =T_{\rm th}\sqrt{1+v\over 1-v },
\end{equation}
for  $v\simeq 0.25c$  and  $T_{\rm th}\simeq 100 $ MeV (a value  generally 
expected for the kinetic freeze-out in the reaction energy domain here considered)
we see that the slope parameter of the spectra would be $T=129$ MeV.
For further details of emerging reference particle spectra  we refer to Ref.\cite{Letessier}. 

It is common to distinguish the pion charge 
polarity in the pion fugacity $\Upsilon_\pi$ using the third  
component of the isospin $I_3$, (or alternatively the pion charge). 
The conventional definition of   $\Upsilon_\pi$ is~\cite{share}:   
\begin{eqnarray}  
\Upsilon_{\pi^+} \equiv  \gamma^{2}_{q}  \lambda_{I3} , \quad  \Upsilon_{\pi^-} \equiv    
\gamma^{2}_{q}  \lambda_{I3}^{-1} , \quad  \Upsilon_{\pi^\pm}\le e^{\beta m_{\pi^\pm}  } .   
\end{eqnarray} 
The last expression reminds us that $\Upsilon_\pi$ is bounded 
by the Bose singularity. The presence of the  
parameter $\gamma_q$ is required since it can influence 
the magnitude of pion charge asymmetry~\cite{Letessier:1999rm}  
when it approaches in magnitude the  Bose singularity value. 
We recall that $\lambda_i$ are particle fugacities  
and    $\lambda_q^3=e^{\mu_{\rm b}/T}$.  
 
From Eq.\,(\ref{pispec1}) we derive by integration over $p_{\rm t}dp_{\rm t}$ the rapidity  
distribution:   
\begin{eqnarray}\label{pispec2} 
{d\tilde N  \over   dy }    = { 2 \pi} {N_\pi}{\int} {{\Upsilon_\pi m_{\rm t} \cosh y  }    
\over {e^{\beta m_{\rm t} \cosh y } - \Upsilon_\pi}}  p_{\rm t} dp_{\rm t} . 
\end{eqnarray} 
We expect matter flow of the system in  the longitudinal direction: 
\begin{eqnarray} 
{{dN(y)} \over {dy}}={\int_{{\eta}_{\rm min}}^{{\eta}_{\rm max}}}{d{\eta}}{{dN(y-\eta)}  \over  {dy}}, 
\quad 
{\beta}_L=\tanh(\eta_{\rm max}), 
\end{eqnarray} 
where ${\eta}_{\rm max}=-{\eta}_{\rm min}$, from symmetry 
about the center of mass, and ${\beta}_L$ is the maximum longitudinal  
velocity. Thus we have the distribution of thermal pions  
\begin{eqnarray} 
{{1} \over {2{\pi}}}{{dN(y)} \over {{p_t}{dp_t}{dy}}}
={N_{\pi}}{\int_{\eta_{\rm min}}^{\eta_{\rm max}}}{{\Upsilon_\pi m_{\rm t} \cosh (y-\eta)  }          
\over {e^{\beta m_{\rm t} \cosh (y-\eta) } - \Upsilon_\pi}}{d\eta}. 
\end{eqnarray} 
Then the rapidity distribution is  
\begin{eqnarray} 
{{dN(y)} \over {dy}}= {2\pi}}{{N_{\pi}} {\int_{{\eta}_{\rm min}}^{{\eta}_{\rm max}}}
\!\!{\int  {  {{\Upsilon_\pi m_{\rm t} \cosh (y-\eta)  }          
\over {e^{\beta m_{\rm t} \cosh (y-\eta) } - \Upsilon_\pi}}        {dp_t}}}{d\eta}. 
\end{eqnarray}  
  

\section{Analysis of the data}\label{analysis}
In order to model of the  charged pion spectra~\cite{Klay}:\\
1) We describe the 
high $p_{\rm t}$ component 
of the pion $m_{\rm t}$ spectra in terms of a thermal Bose  
shape, which in fact in this case is essentially an exponential. In this 
way we obtain the effective inverse slope $T$. This $T$ comprises 
a combination of the intrinsic thermal temperature $T_{\rm th}$ and the 
shift due to  the collective longitudinal and radial expansion flow. 
The longitudinal flow is required for an accurate description of the 
rapidity spectra.\\
2)  We form the difference between the direct pion spectrum 
determined by the high $m_{\rm t}$ fit and the 
experimental spectrum. We then model this difference in terms 
of  the pion  decay spectrum $\Delta_{33}(1232)\to N+\pi$. We form the 
pion spectrum from direct and decay components using 
a fitted strength of the not-scattered $\Delta_{33}$-decay 
component. \\
3) We find that 
the pion spectrum can be described precisely when we 
reduce all hadron masses  using a uniform multiplicative factor. A further 
improvement is apparently possible when  the radial flow is introduced, 
but we do not pursue  such refinements here for reasons 
already discussed. 

The  thermal  momentum distribution of  primary pions 
is, by assumption,  given  by the thermal Bose 
distribution Eq.\,(\ref{pispec1}). In the center of the hot, 
dense collision zone, high pressure is produced.  This pressure 
will cause a collective motion of the system which has 
been experimentally observed~\cite{Lisa:1994yr,Siemens}. However, it 
is not possible for us to  disentangle thermal and collective flow 
effects~\cite{Brown:1991en}.  Thus, the $m_\bot$-slope we obtain 
comprises some Doppler-like shift due to source motion.  
Aside of the collective radial motion we expect that, in  the longitudinal direction, 
the memory of the projectile-target motion will remain. 

Such longitudinal flow ${\beta}_L$ is considered in Ref.\,\cite{Klay}, employing a 
flat longitudinal rapidity profile bounded by  ${\eta}_{max}=-{\eta}_{min}$, 
where ${\beta}_L=\tanh(\eta_{max})$. A somewhat simpler 
approach is to consider two fluids  moving apart with a 
remainder of the original projectile and target rapidity:
\begin{eqnarray}\label{longflow}
{{d^2\!N(p_{\rm t},y)} \over {p_{\rm t} dp_{\rm t}dy}}
 ={{1} \over {2}}{{d^2\!\tilde N(p_{\rm t},y-\eta_0)}  
\over  {p_{\rm t} dp_{\rm t}dy}} \!+\!   {{1} \over {2}} {{d^2\!\tilde N(p_{\rm t},y+\eta_0)}  
\over  {p_{\rm t} dp_{\rm t}dy}}
\end{eqnarray} 
We will fit $\eta_0$ to the experimental rapidity distributions. 
For the $m_{\rm t}$ spectrum inclusion of these two sources means 
that   at $y=0$ we use Eq.\,(\ref{pispec1}) substituting $m_{\rm t}\to {m_{\rm t}}{\cosh {\eta_0}}$.

\begin{figure*}[thb]
\centerline{\psfig{width=13.8cm,figure= 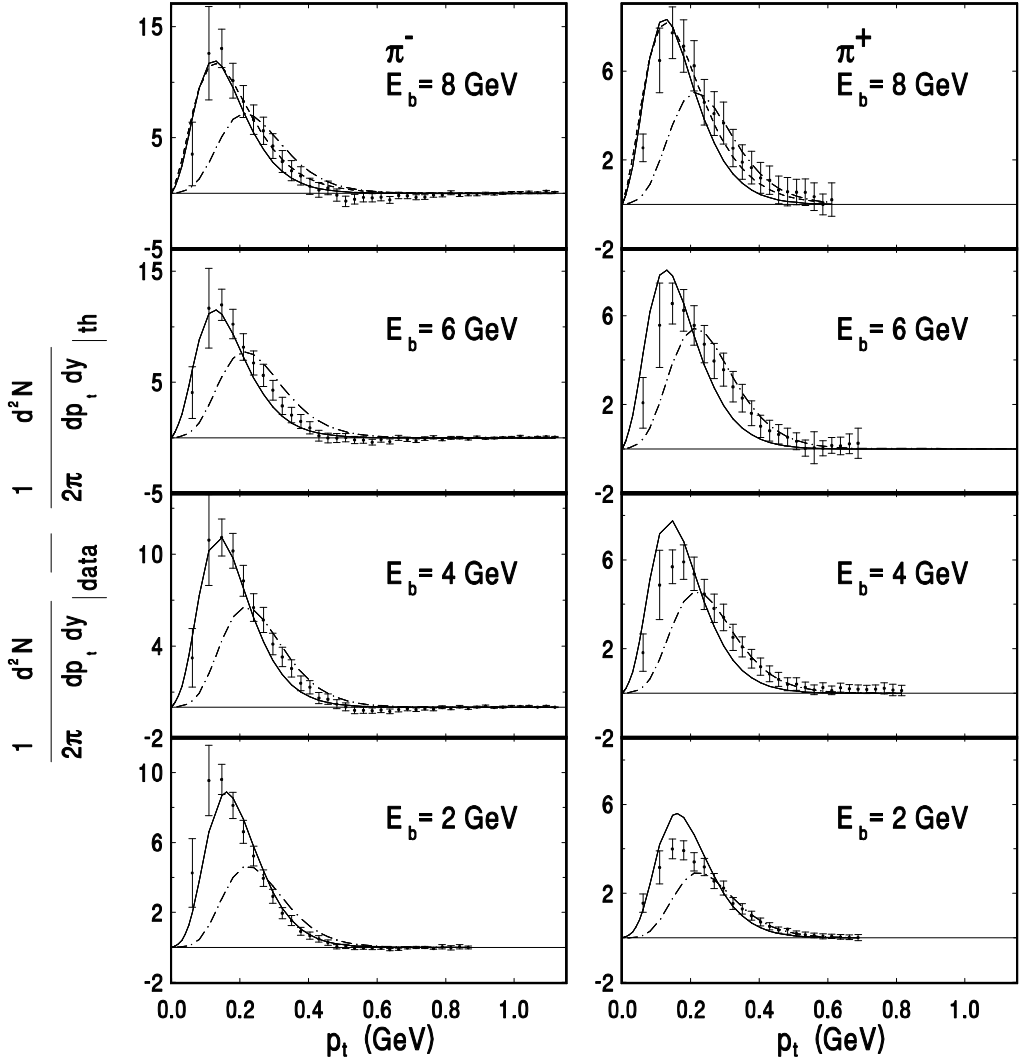}}
\caption{\label{Difference}Difference of experimental $\pi^\pm$ yields and the   direct thermal  (th)
yields, for the mid-rapidity data bin~\cite{Klay}. The contribution from 
mass modified $\Delta_{33}$ decay is shown by solid line.
 Contribution from mass modified $\Delta_{33}$ decay  with radial 
flow is shown by dashed line for the case of 8 GeV 
beam energy (top panels). Dot-dashed line shows the $\Delta_{33}$ decay 
contribution without mass modification.}
\end{figure*}

The parameters of the thermal model   are presented   
in the top section of table \ref{BRtable1},
 where $T_{\rm slope}= T/{\cosh {\eta_0}}$. 
$T$ comprises the effect of transverse expansion dynamics.
For large $p_{\rm t}$ the 
pion fugacity is irrelevant:  the thermal spectrum has a Boltzmann  
shape and is dominated by direct thermal component, 
and the fugacity becomes another component in 
the yield normalization. In this limit  value of the mass 
of the pion is also irrelevant, as it is negligible compared to the
value of $p_\bot$. Hence  the domain of large $p_{\rm t}>> m_\pi$  fixes the  value of 
$T_{\rm slope}$.
There remains a  major difference between experimental spectrum  and a single component thermal 
spectrum  as was noted in \cite{Klay}. This difference is shown  in Fig.\,\ref{Difference}.
This difference   appears  similar 
at all reaction energies and  is localized in to a momentum 
range $p_{\rm t}<0.4$ GeV. This suggests as the common
and unaccounted mechanism of pion 
production  the $\Delta_{33}$-resonance decay.   

\begin{table}[htb]
\caption{Top: statistical parameters used to describe $\pi^\pm$ spectra; bottom: 
modified hadron masses as used at different beam energies.
\label{BRtable1} }
\begin{tabular}{|l|c|c|c|c|}
\hline
Beam Energy  &2$A$ GeV  &4$A$ GeV    &6$A$ GeV    &8$A$ GeV       \\
\hline
$T_{\rm slope} $  [MeV]         &       121    &     128    &      133   &       135    \\
$\eta_{0}$                      &       0.15   &    0.25    &      0.3   &       0.3    \\
$T    $ [MeV]                  &        122    &     132    &      139   &       141    \\                                       
$\lambda_{I3}^{-1}$             &      1.2     &    1.15    &      1.15  &       1.15   \\
$\gamma^{2}_{q}$                &      1.85    &    1.62    &      1.44  &       1.44   \\
$N_{\pi^{-}}[{\rm GeV}^{-3}]$   &       120    &     339    &      521   &       647    \\
$N_{\pi^{+}}[{\rm GeV}^{-3}]$   &       136    &     383    &      587   &       727    \\ 
$N_{\Delta}e^{\mu_{\rm B}/T_{\rm ch}}  $
  &     2,170    &     983    &      630   &       536    \\
$m/m_0            $             &      0.72     &    0.63    &   0.59     &      0.57    \\
\hline
$m_{\Delta}[{\rm GeV}]$         &       0.880  &    0.774   &    0.722   &      0.704   \\
$m_{\rm N}[{\rm GeV}]$          &       0.671  &    0.590   &    0.550   &      0.537   \\
$m_{\pi}[{\rm GeV}]$            &       0.100  &    0.088   &   0.082    &      0.080   \\
\hline
${\chi^2}/ $ dof                &       1.16   &    1.20    &     1.04   &      1.19    \\
\hline
\end{tabular}
\end{table} 

For small $p_{\rm t}$ there is considerable impact of the Bose 
nature of the pion on the spectra. Furthermore, the difference in the shape between 
the positive and negative pion spectra can be in part 
accommodated by differences in $\Upsilon_{\pi^\pm}$, and in part by the 
associated difference in the relative yields of the $\Delta_{33}$-resonances 
as indicated  by the power of $\lambda_{I3}$ for each 
of the polarities, see last column in table \ref{BRtable2}. 
The branching into the different channels for each of the $\Delta_{33}$-resonances 
is relevant considering the isospin weights. 
Since we allowed
for the isospin asymmetry in the fugacity, we could insist in normalizing the $\pi^\pm$ 
yields by same common factor. On the other hand, a considerable 
improvement of the description of the spectra arises if the 
normalizations differ slightly, within the 15\% uncertainty of 
the experimental data.  The normalization factors we use are also 
presented   in table  \ref{BRtable1}. Aside from experimental effects, such 
minor variation could  also arise from differences in rescattering 
cross sections of charged pions in neutron-rich baryonic matter.

\begin{table}[tbh]
\caption{Delta decays and fugacities. 
\label{BRtable2}}
\begin{center}
\begin{tabular}{|lcc|l|}  
\hline
              &        & Decay          Channel                    &       
\hspace*{0.21cm}       
$\Upsilon_\Delta$\\\hline$\Delta^{++}$ & $\to $ & $   (\pi^+  + p )    $ & $  \gamma^{3}_{q} \lambda_{I3}^{3/2}$ \\
$\Delta^{+} $ & $\to $ & $   1/3 (\pi^+  + n),\  2/3 (\pi^0 + p )$ & $  \gamma^{3}_{q} \lambda_{I3}^{1/2}$ \\
$\Delta^{0} $ & $\to $ & $   1/3 (\pi^-  + p ),\  2/3 (\pi^0 + n) $ & $  \gamma^{3}_{q} \lambda_{I3}^{-1/2}$ \\ 
$ \Delta^{-}$ & $\to $ & $      (\pi^-  + n)                     $ & $   \gamma^{3}_{q} \lambda_{I3}^{-3/2}$  \\
\hline
\end{tabular}
\end{center}
\end{table}

To describe this difference spectrum quantitatively, we consider 
$\Delta_{33}$ two body decay,  and employ the well known 
results, see, {\it e.g.\/}, Ref.\cite{Sollfrank:1990qz,Sollfrank:1991xm}.  
\begin{eqnarray}
{{d\tilde N_{de}} \over {{p_t}{dp_t}{dy}}}
&=&{{{m_{R}}b}  \over  {4{\pi}{p^{*}} }}{\int_{y_{R}^{(-)}}^{y_{R}^{(+)}}}{{dy_{R}} 
\over  {\sqrt{{m_{t}^2}{{\cosh}^2}(y-y_R)-{p_t^2}}}}  \times  \nonumber  \\
&& \hspace*{-2.25cm}
\times {\int_{m_{tR}^{2(-)}}^{m_{tR}^{2(+)}}}\!\!\!\!{{dm^2_{tR}}   \over 
{\sqrt{(m^{(+)}_{tR}-m_{tR})(m_{tR}-m_{tR}^{(-)})}}}{{d\tilde N_{R}}  
\over  {{m_{tR}}{dm_{tR}}{dy_R}}},
\end{eqnarray}
where $b$ is the branching ratio, and $p^{*}$ is the momentum 
of the decay particle in the $\Delta_{33}$-rest-frame. For 
the $\Delta_{33}$-resonance spectrum we assume at first a Boltzmann   
distribution,  with $T$ determined by  the pion 
spectrum and  the $\Delta_{33}$-fugacity shown in table \ref{BRtable2}. 
\begin{eqnarray}\label{DeltaY}
{{1} \over  {2\pi}}{{d\tilde N_R} \over {{m_{tR}}{dm_{tR}}{dy}}}
={N_{\Delta}}e^{\mu_{\rm B}/T_{\rm ch}}       {{ {\Upsilon_{\Delta}} {m_{tR}}{\cosh {y_R} }}     
\over {\exp({{\beta}{m_{tR}}{\cosh {y_R}}})}}.
\end{eqnarray}
In the resulting pion decay component we  also introduce  
the longitudinal flow according to Eq.\,(\ref{longflow}). In this way 
we allow effectively for the longitudinal motion of the $\Delta_{33}$-source. 
The total pion distribution is composed of the direct and the 
decay  contributions.

When computed with the  free space  masses of $\pi,\Delta_{33},N$  
 the resulting spectrum of decay pions is found to have the 
shape seen in  Fig.\,\ref{Difference}, but is  systematically shifted to   
a higher  momentum than the experimental data. In order to 
reduce the energy scale to the level seen in the  difference spectrum  
in Fig.\,\ref{Difference}, we scale down the three hadron   
masses  $\pi,\Delta_{33},N$ by the  same factor, which is chosen 
in a qualitative manner. The resulting masses we use are stated in 
the bottom section of table \ref{BRtable1}; the modification is non-negligible. There 
is a tendency for the mass reduction effect to increase with collision 
energy. As is seen in Fig.\,\ref{Difference} (solid lines), after this modification  we describe 
the experimental data rather well.

\begin{figure*}[thb]
\centerline{\psfig{width=13.3cm,figure= 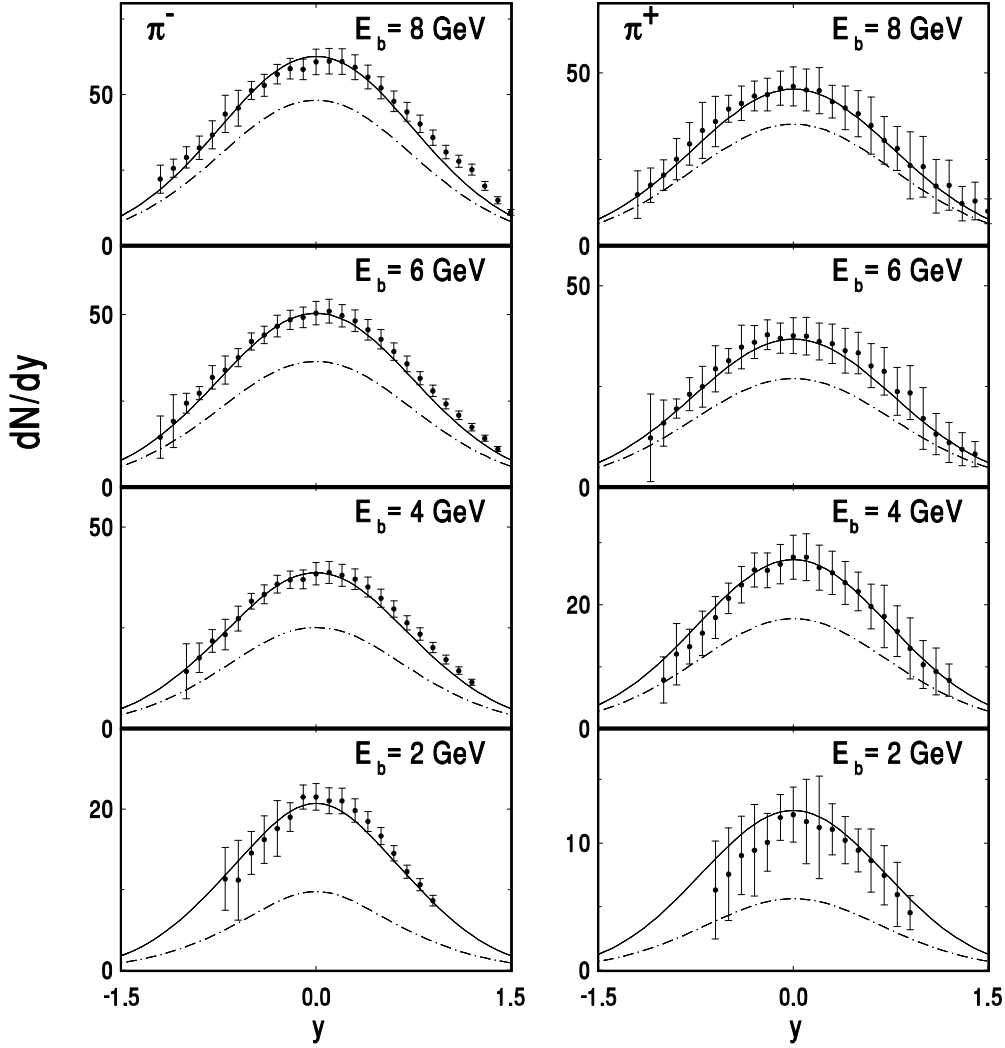}}
\caption{\label{Rapidity}$\pi^\pm$ rapidity distribution~\cite{Klay}, compared with  model (see text). Dot-dashed: 
no $\Delta_{33}$-decay pion contribution. }
\end{figure*}

We also have to allow for the presence of matter flow. 
In general  the radial flow has  the   effect of  flattening the $\Delta_{33}$ 
spectrum more than the $\pi$ spectrum, since the effective 
inverse slope $T$, which combines the intrinsic temperature  
$T_{\rm th}$ with the blue-shift of flow, grows somewhat with particle 
mass   and collective velocity $ v$~\cite{Bearden:1996dd}.
We account for the difference that
the flow effect has on $\pi$, and the much heavier $\Delta_{33}$ as follows: 
We   keep all  quantities except for the distribution $N_{\Delta}$ 
unchanged. We  modify the   $\Delta_{33}$-spectrum 
using a simplification of the model proposed in Ref.\,\cite{Dobler:1999ju}. 
For the flow rapidity profile  $\eta_{\rm t}(r)=\eta_{\rm tf}(r/R)^{\alpha}$,
we take $\alpha =0$ and $\eta_{tf}=0.3$. We allow here a cylindrical 
fireball   since this is not a significant element in our approach 
and the model is simple and transparent.    
The contributions from the $\Delta_{33}$ decay with radial flow are 
shown in Fig.\,\ref{Difference} solely for 8 GeV beam energy,
where this effect can be expected to be largest, 
as dashed lines.   There is a mild improvement in the high $p_{\rm t}$ part 
of the difference spectrum. These results suggest  that the complete and detailed 
inclusion of the  radial flow, 
which has the largest influence for highest reaction energy, will further
improve the understanding of the difference spectra 
seen in  Fig.\,\ref{Difference}.

The  integral of the transverse momentum spectra leads to the rapidity distribution   shown 
in Fig.\,\ref{Rapidity}. In performing this integral it is relevant for the rapidity dependent 
normalization to disentangle in the slope temperature the flow effect from the true temperature.
Since we already have seen that matter flow is desired, it is not 
surprising that when we do this introducing  for simplicity a longitudinal 
flow $\eta_0$, the rapidity spectrum is described very well. 
The resulting $\chi^2$/dof is given in the bottom of 
table \ref{BRtable1}. It should be remembered that this 
is not a best fit optimizing the 8 parameters (4 statistical, 3 normalization 
parameters, and one mass scale parameter which determines the three hadron masses) 
at each reaction energy, and that we have treated the flow in somewhat cavalier, but
as we think, precise enough manner.

\section{Discussion and Summary}\label{sum}

We have obtained an accurate description of pion 
$\pi^\pm$ spectra, consistent across the 
four reaction energies considered here. Our
objective we had here was to show that the pion spectra we consider 
are sensitive to the introduction of the mass change 
parameter. We believe that we have  
demonstrated this in  that we can explain quite satisfactorily 
the spectra considered. The relevant model parameters 
were  determined in part by using the common knowledge and in 
 part optimizing the parameters to  the data. We did not make an effort to 
minimize the value of $\chi^2$, since several important physics 
elements are beyond the scope of 
the current approach. 

We generally 
proceeded as follows: we chose  a reasonable value of  
the hadron mass reduction parameter, the longitudinal 
flow $\eta_0$, and the two fugacities. Once this choice 
is made, the temperature $T$ (common for both $\pi^+$ and $\pi^-$),  the  two 
pion normalization factors $N_{\pi^\pm}$ and the $\Delta_{33}$-yield 
normalization are fitted to the data. We then tried another reasonable 
set of the three initial parameters in an effort to further  
reduce the $\chi^2$ and did this several times.

We would like to make a few additional remarks 
about the physical meaning of the parameters, possible 
future directions, and consistency of our approach:\\
1) One could argue that the softening of the pion decay spectrum,
 {\it i.e.\/}, the difference spectrum seen in Fig.\,\ref{Difference} is not due 
to mass modification but is due to decay pion scattering in matter. 
It is however hard to understand why this process at all reaction energies would have the 
effect  of producing  the $\Delta_{33}$-decay spectrum 
shape, with reduced energy scale. We note that though 
the data description we present is relatively good, at  low $p_{\rm t}$
 there is some potential to improve the model, e.g. by considering the effect of pion scattering. \\
2) In table \ref{BRtable1},  $N_{\Delta}$ is common for different 
isospin states of $\Delta$. For 6, 8 GeV beam energy 
cases,   $N_{\Delta}$ and $N_{\pi}$  are comparable. However,  
 $N_{\Delta}>N_{\pi}$ for 2 GeV beam energy. This suggests 
that the direct thermal pion component is dominated by the 
indirect pion production via the $\Delta_{33}$-resonance. Indeed, as 
is seen in Eq.\,(\ref{DeltaY}), the $\Delta_{33}$-yield   comprises a  
factor, $e^{\mu_{\rm B}/T_{\rm ch}}$, where $\mu_{\rm B}$ is the baryon 
chemical potential. The rapid growth in the  normalization  factor 
with  decreasing reaction energy is due to the expected 
increase of $\mu_{\rm B}/T_{\rm ch}$. \\
3) Hadron model calculations~\cite{Rapp} find mass shifts which are small, and
decay width  increases  which are large.  We note that the effect obtained in 
the bag model effect differs. The decrease in the hadron masses 
is accompanied by a decrease in transition energy, and thus
the decay phase space diminishes. This reduces the width. There is a further potential 
reduction due to the reduction of the intrinsic decay matrix element, since for a smaller value of the
bag constant $B $ the quark bag grows in size, 
diluting the strength of the reaction matrix element. We recall that at present an
increase of mass shift has not been observed experimentally, while a 
decrease in width is not inconsistent with the available data. Thus a priori, a 
bag model description seems to agree better with data. \\   
4) The consistency of our model can be tested by introducing the 
actual $\Delta_{33}$-resonance spectra obtained from a reconstruction 
by the invariant mass method  of  the mass 
modified $\Delta_{33}$-resonance using modified masses of decay 
pions and nucleons. Success of this would indeed be a 
convincing step supporting the path of analysis proposed here. 
Also, it would allow us  to   include in our approach the 
spectra of nucleons.\\ 
5) The  matter flow field arises from initial conditions and ensuing dynamics 
driven by equations of state. Thus there is in principle a relation between
the magnitude of the flow and the freeze-out condition (temperature, matter density) for each 
heavy ion reaction energy, if initial conditions could be understood. Our 
 results are consistent with what may
be called, in absence of theoretical understanding, the general wisdom 
about the magnitude of matter flow, temperature and density 
expected in conditions   considered here. \\

In summary, we have studied the precision spectra of $\pi^\pm$ obtained for kinetic beam energy 
per nucleon 2,\,4,\,6,\,8 $A$ GeV at AGS. We can interpret the $\pi^\pm$ spectra as  originating in a thermal fireball of 
dense matter. The necessary ingredients are aside of direct $\pi^\pm$  production, 
secondary $\Delta_{33}$-resonance decay, longitudinal 
flow, and, a significant   mass modification  of the involved hadrons. 
In fact, the mass modification is the  new physics 
element in the   analysis presented,  allowing  a significant improvement 
of the data description using a single thermal 
source.

\begin{acknowledgement}
JR would like to thank PZB for very kind hospitality at the Fudan 
University where this work was initiated. PZB was 
supported in part by Fudan University under contract EX13314,  
sand JR was  supported by a grant from the U.S. Department of 
Energy, DE-FG02-04ER41318.
\end{acknowledgement}
\vspace*{0.3cm}

\end{document}